\documentclass[useAMS,usenatbib,letters]{mn2e}
\usepackage{epsf}

\title[Gravitational lens time delays for distant supernovae]
{Gravitational lens time delays for distant supernovae: break the
degeneracy between radial mass profiles and the Hubble constant}
\author[M. Oguri and Y. Kawano]
{Masamune Oguri$^{1}$\thanks{E-mail:
oguri@utap.phys.s.u-tokyo.ac.jp} and Yozo
Kawano$^{2}\thanks{E-mail:
kawano@a.phys.nagoya-u.ac.jp}$\\
$^{1}$Department of Physics, School of Science, University of Tokyo,
Tokyo 113-0033, Japan\\
$^{2}$Department of Physics and Astrophysics, Nagoya
University, Nagoya 464-8062, Japan}

\voffset- .65in

\begin{document}

\date{\today}

\pagerange{L\pageref{firstpage}--L\pageref{lastpage}} \pubyear{2002}

\maketitle

\label{firstpage}

\begin{abstract}
An attempt to measure the Hubble constant with gravitational lens time
delays is often limited by the strong degeneracy between radial mass
profiles of lens galaxies and the Hubble constant. We show that 
strong gravitational lensing of type Ia supernovae breaks this
degeneracy; the standard candle nature of type Ia supernova luminosity
function allows one to measure the magnification factor directly, and
this information is essential to constrain radial mass profiles and the
Hubble constant separately. Our numerical simulation demonstrates that
the Hubble constant can be determined with $\sim 5$\% accuracy from only
several lens events if magnification factors are used as constraints.
Therefore, distant supernova survey is a promising way to measure the
global Hubble constant independently with the local estimates. 
\end{abstract}


\section{Introduction}\label{sec:intro}

Gravitational lensing is known to be a powerful tool to determine the
Hubble constant $H_0$ directly, without using the distance ladder
\citep{refsdal64}. Derived values of $H_0$ show good agreement
among known several lens systems, once the radial mass
profiles of lens galaxies are fixed \citep{koopmans99,kochanek02a}.
If the lens galaxies are assumed to have the singular isothermal mass
distribution, analysis of five gravitational lens systems indicates that
the value of $H_0$ is $H_0\sim 50{\rm km\,s^{-1}Mpc^{-1}}$
\citep{kochanek02a} and hence too low to be consistent with the local
measurement $H_0\sim 70{\rm km\,s^{-1}Mpc^{-1}}$ \citep{freedman01}. 
This discrepancy may be ascribe to inhomogeneity in the universe
\citep{tomita00a,tomita00b}, and therefore it is important to study the
global Hubble constant independently with the local
measurement which relies on the distance ladder.

The main limitation of this technique is that there is a strong
degeneracy between radial mass profiles and $H_0$
\citep*{wambsganss94,keeton97,koopmans99,witt00,tada00,williams00,chiba02,wucknitz02,kochanek02a,zhao03}.
Therefore, unless we specify the radial mass distribution in the lens object,
we hardly constrain the value of $H_0$. The strong dependence of
differential time delays on the radial mass distribution, on the other
hand, indicates that statistics of time delays provide a powerful probe
of density profiles \citep{oguri02}. This degeneracy can be broken if the
Einstein ring images of host galaxies are observed \citep*{kochanek01},
but the observation of host galaxies is often very difficult because of
the large brightness contrast between quasars and host galaxies. Other
way to break this degeneracy comes from the central core images
\citep{rusin01,evans02,keeton03}. The lack of central core images, however,
places only the lower limit of the mass concentration, and this corresponds
to the lower limit of $H_0$. Information of stellar kinematics and
the mass-to-light ratio also allows one to constrain the radial mass profile of
the lens galaxy and to break the degeneracy in the Hubble constant,
although the measurement of velocity dispersions for distant galaxies is
very difficult and involves large uncertainties. \citet{treu02}, for
instance, concluded that the mass density profile of the lens PG1115+080
is steeper than the singular isothermal mass distribution, and obtained
a value of the Hubble constant  $H_0\sim 60{\rm km\,s^{-1}Mpc^{-1}}$
which is marginally consistent with the local measurement. 

In this {\it Letter}, we consider strong lensing of distant supernovae
(SNe) which is expected to be observed by future observational plans 
\citep*{kolatt98,wang00,porciani00,sullivan00,holz01,goobar02,oguri03}. 
For instance, the satellite {\it SNAP}\footnote{http://snap.lbl.gov/}
 (SuperNova/Acceleration Probe) will catch at least a few lensed SNe per year. 
The well-known advantages of gravitational lensing for distant supernovae
include: (1) differential time delays between images can be easily
measured with high accuracy (of order of one hour), in marked contrast
with usual quasar lensing for which time delay measurement is quite
difficult \citep[e.g.,][]{kundic97}; (2) since SNe are transient
phenomena, it is easier to observe lensed host galaxies of SNe after SNe
are faded away. More importantly, we find that information of
magnification factors, which can be directly measured in the case of
type Ia SN lensing, breaks the degeneracy between radial mass profiles
and $H_0$. First we provide the reason for this degeneracy breaking
analytically, and next we numerically demonstrate the importance of
magnification factors. In what follows, we adopt a lambda-dominated
universe with $(\Omega_0,\lambda_0)=(0.3,0.7)$, where $\Omega_0$ is the
density parameter and $\lambda_0$ is the cosmological constant. The
Hubble constant in units of $100{\rm km\,s^{-1}Mpc^{-1}}$ is denoted by $h$.

\section{Analytic Consideration}\label{sec:ana}

In this section, we estimate how accurately we can constrain radial mass
profiles and the Hubble constant separately, on the basis of simple
analytic consideration. Following \citet{wucknitz02}, we consider the
degeneracy between radial mass profiles and the Hubble constant in terms
of well-known mass-sheet degeneracy \citep*{falco85}. First start from
the lens equation:
\begin{equation}
 \bmath{y}=\bmath{x}_i-\bmath{\nabla}\psi(\bmath{x}_i),
\end{equation}
where $\bmath{y}$ and $\bmath{x}_i$ are the positions of the source and images,
and $\psi(\bmath{x}_i)$ denotes the lens potential \citep*[see][]{schneider92}.
If we multiply the lens equation with $(1-\kappa)$, then the lens
equation can be rewritten as
\begin{equation}
 (1-\kappa)\bmath{y}=\bmath{x}_i-\bmath{\nabla}\left[(1-\kappa)\psi(\bmath{x}_i)+\kappa\frac{x_i^2}{2}\right].
\end{equation}
Therefore, the image position $\bmath{x}_i$ is never changed if we transform
{\it unobservable} quantities as $\bmath{y} \rightarrow (1-\kappa)\bmath{y}$
and $\psi(\bmath{x}_i)\rightarrow (1-\kappa)\psi(\bmath{x}_i)+\kappa(x_i^2/2)$.
This is the mass-sheet degeneracy. Since the time delay is calculated from
\begin{equation}
 h\Delta t_i \propto \frac{(\bmath{y}-\bmath{x}_i)^2}{2}-\psi(\bmath{x}_i),
\end{equation}
this transform also changes the estimation of the Hubble constant, $h
\rightarrow (1-\kappa)h$ if $\Delta t_{ij}$ is fixed to the observed value. 
Therefore, the Hubble constant $h$ cannot be uniquely determined from
information of $\{\bmath{x_i}\}$ and $\{\Delta t_{ij}\}=\{\Delta t_i-\Delta
t_j\}$. In the usual quasar lensing, additional information is supplied
by the flux ratio. The flux ratio is simply derived from the ratio of
magnification factors $\mu_i$:
\begin{equation}
 \mu_i=\left|\frac{\partial \bmath{y}}{\partial \bmath{x}_i}\right|^{-1}.
\end{equation}
From this expression, it is found that $\mu_i$ is transformed as $\mu_i
\rightarrow (1-\kappa)^{-2}\mu_i$. Hence the flux ratio
$r_{ij}=\mu_i/\mu_j$ is also never changed. This means that information
of $r_{ij}$ cannot break this degeneracy. The degeneracy between radial
mass profiles, $\psi\propto r^\beta$, and the Hubble constant $h$ can be
interpreted in this context because of the simple relation
$1-\kappa=2-\beta$ \citep{wucknitz02}. This yields a general scaling
law, 
\begin{equation}
 h\propto 2-\beta,
\label{deg_h}
\end{equation}
without changing observable values such as $\bmath{x}_i$, $\Delta
t_{ij}$, and $r_{ij}$.

In the case of SN Ia lensing, the situation changes drastically.
``Standard candle'' nature of SNe  Ia \citep*{phillips93,riess96} allows
one to observe magnification factors $\mu_i$ directly. The
magnification factor has the strong dependence on the radial mass
profile (see also \citealt{wambsganss94}; Kawano et al., 
in preparation),
\begin{equation}
 \mu_i \propto (2-\beta)^{-2},
\label{deg_mu}
\end{equation}
thus we can break the degeneracy by the measurement of magnification
factors. We note that \citet{kolatt98} also proposed this method to break
the mass-sheet degeneracy in galaxy clusters and reconstruct galaxy
cluster mass. Although $\mu_i$ should have significant error which
arises from the intrinsic dispersion in SNe Ia peak luminosities as well as
substructure in the lens galaxy \citep{mao98}, its effect on $h$
estimation is not so severe because equations (\ref{deg_h}) and
(\ref{deg_mu}) implies that the resulting error of $h$ is a half of that
of $\mu_i$.
 
\section{Simulated Result}\label{sec:sim}

To illustrate how accurately we can determine $\beta$ and $h$, we show
the results of our simulation. First, we assume the lens galaxy is well
characterized by the following lens potential (Kawano et al., in
preparation):
\begin{equation}
 \psi(\bmath{x})=r^\beta(a_0+a_2\cos 2\theta+a_3\cos 3\theta+b_2\sin 2\theta+b_3\sin 3\theta),
 \label{lenspot}
\end{equation}
where $\bmath{x}=(r,\theta)$ is the position in the polar coordinate. We note
that best-fit values of $h$ and $\mu$ in the analysis of quasar
PG1115+080 are quite insensitive to the choice of the angular part of
the lens potential (\citealt{kochanek02b}; Kawano et al., in preparation), 
thus it is sufficient to analyze this lens potential only. We also take 
account of the effect of external shear as 
\begin{equation}
 \psi_{\rm shear}(\bmath{x})=\frac{1}{2}\gamma r^2\cos(2\theta-2\theta_\gamma).
\end{equation} 
We randomly put a source, generate quadruple images, and calculate
differential time delays and magnification factors, assuming following
parameters: 
$\beta=1.0$, $h=0.5$, $a_0=0.5$, $a_2=b_2=0.01$, $a_3=b_3=0.001$,
$\gamma_1=\gamma\cos 2\theta_\gamma=0.1$, and
$\gamma_2=\gamma\sin 2\theta_\gamma=-0.01$, where $a_i$ and $b_i$ are
values when $\bmath{x}$ is in units of arcsec. 
Source and lens are placed at $z_{\rm S}=1.5$ and $z_{\rm L}=0.5$,
respectively. Generated images have separations on the order of $1''$. 
For the observable quantities, such as image positions and
time delays, the Gaussian noise is added. We assume following dispersions: 
\begin{eqnarray}
 \sigma_x&=&0.01'',\\
 \sigma_{\Delta t}&=&0.05\,{\rm [day]},\label{dis_td}\\
 \sigma_{\log r_{ij}}&=&0.09,\\
 \sigma_{\log \mu_i}&=&0.12.
\end{eqnarray}
The precisions in positions and time delays are consistent with 0.1
pixel of the instrument and estimated accuracy for SNe Ia lightcurves
in SNAP survey, respectively \citep{goobar02}. The dispersion of
magnification ratio, which roughly corresponds to $\sim 20$\% fractional
error, is a fiducial error often assumed in $\chi^2$ minimization
\citep[e.g.,][]{kochanek02a}. We assume that the dispersion of the
magnification factor is somewhat larger than this, roughly corresponds
to $\sim 30$\% fractional error, because not only substructure in the
lens galaxy \citep{mao98} but also the intrinsic dispersion of SNe Ia
peak magnitudes contribute to $\sigma_{\log\mu_i}$. Other possible
source of the dispersion is dust extinction in the lens galaxy. However,
the effect of dust extinction can be corrected from the observed
reddening because of knowledge of an SN Ia's intrinsic color
\citep[e.g.,][]{riess96}. 

\begin{figure}
  \begin{center}
    \epsfxsize=8.1cm
    \epsfbox{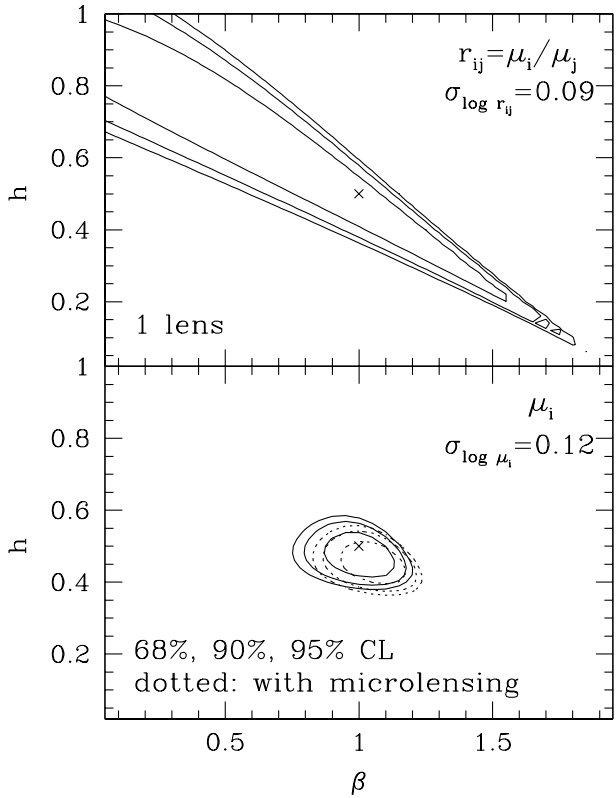}
  \end{center}
    \caption{Constraints on the radial mass profile $\beta$ (eq.
 [\ref{lenspot}]) and the Hubble constant $h$. The contours of
 $\Delta\chi^2=\chi^2-\chi_{\rm min}^2$ in the $\beta$-$h$ plane are
 calculated from one quadruple lens event. Crosses indicate the assumed
 value in generating observable quantities; $(\beta,h)=(1.0,0.5)$. {\it Upper
 panel}: image  positions $\bmath{x}_i$, differential time delays 
 $\Delta t_{ij}$,  and magnification ratios $r_{ij}$ are used to
 calculate $\chi^2$. {\it Lower panel}: instead of $r_{ij}$,
 magnification factors $\mu_i$ are used to calculate $\chi^2$. Dotted
 lines are same as solid lines, but in this case additional non-Gaussian
 errors due to microlensing are also included.} 
\label{fig:betah}
\end{figure}
\begin{figure}
  \begin{center}
    \epsfxsize=8.1cm
    \epsfbox{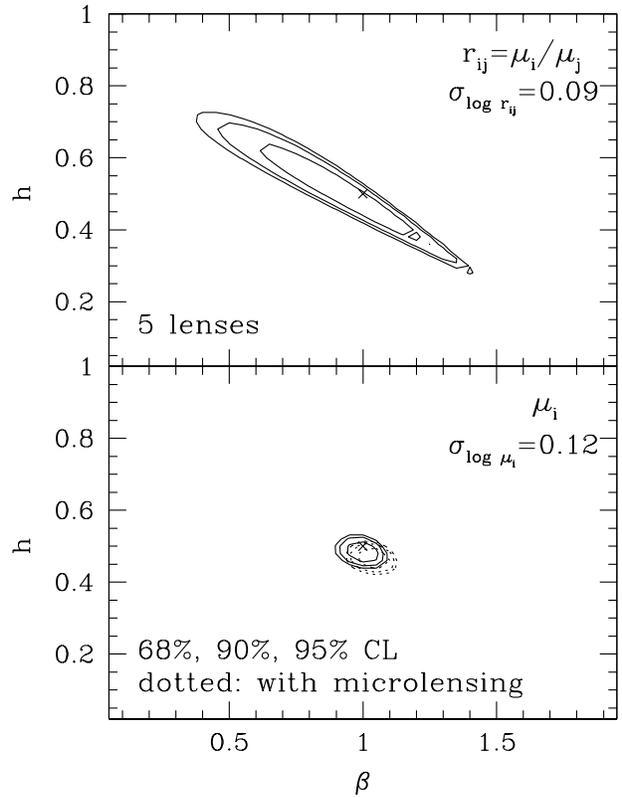}
  \end{center}
    \caption{Same as Figure \ref{fig:betah}, but from five quadruple lens
 events. } 
\label{fig:betah_comb}
\end{figure}

\begin{figure}
  \begin{center}
    \epsfxsize=8.1cm
    \epsfbox{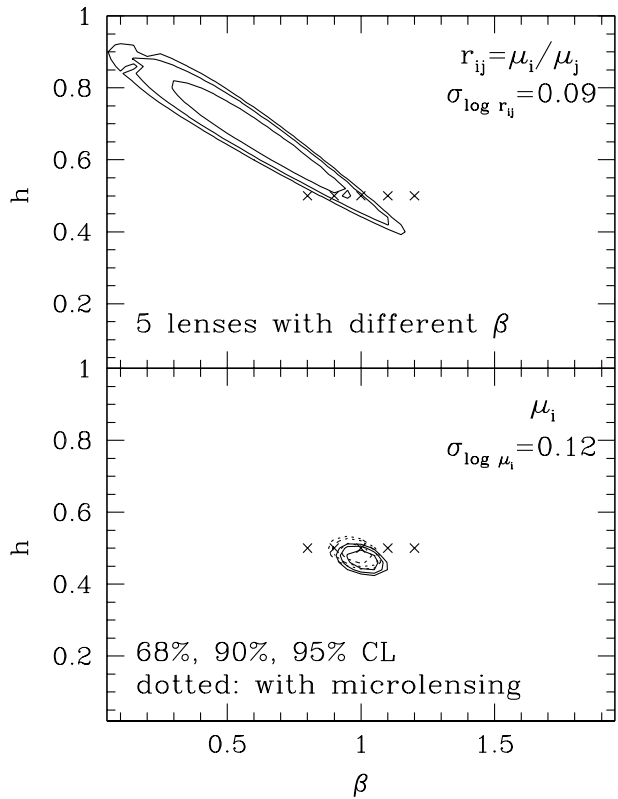}
  \end{center}
    \caption{Same as Figure \ref{fig:betah}, but from five quadruple lens
 events with different radial mass profiles; $\beta=0.8$, $0.9$, $1.0$, 
 $1.1$, and $1.2$. }  
\label{fig:betah_comb_difbeta}
\end{figure}

After the virtual ``observational data'' is generated, we perform
$\chi^2$ minimization using the same lens model. At that time, we fix
values of $\beta$ and $h$, and optimize the other parameters such as
$a_i$, $b_i$, $\gamma$, and the source position. Faint core images which
may appear when $\beta>1$ are always neglected. In calculating $\chi^2$,
we consider following two cases: (1) Only the magnification ratio $r$ is
measured. This case corresponds to traditional quasar lensing. (2) The
magnification factor is directly measured. This is the case of SN Ia
lensing we are interested in. For each case, we calculate the contour of
$\Delta\chi^2=\chi^2-\chi_{\rm min}^2$ in the $\beta$-$h$ plane. Figure
\ref{fig:betah} plots constraints on $\beta$ and $h$ from one quadruple
lens event. This figure clearly shows that in the case of SN Ia lensing
$\beta$ and $h$ are well constrained separately. It is surprising that
the Hubble constant is determined with $\sim 10$\% accuracy (68\%
confidence) from only one lens system. On the other hand, when
magnification factors are not used, $\beta$ and $h$ are poorly
determined; they show the strong degeneracy $h\propto 2-\beta$. We note
that in practice constraints from quasar lensing may be worse than
our result using magnification ratios, because the error of time delays
is usually much larger than our assumption (eq. [\ref{dis_td}]). Figure
\ref{fig:betah_comb} shows constraints from five quadruple lens events.
In generating observables for each event, the position of the source is
changed while the lens model is always fixed. In this figure, the Hubble
constant $h$ is determined with $\sim 5$\% accuracy (68\% confidence)
when magnification factors are used, while the accuracy is still $\sim
20$\% (68\% confidence) when magnification ratios are used. We also
examine the case that lens galaxies have different values of $\beta$,
and the result is shown in Figure \ref{fig:betah_comb_difbeta}. In this
plot, we assume that five lens systems have different radial mass
profiles; $\beta=0.8$, $0.9$, $1.0$, $1.1$, and $1.2$, respectively.
This figure clearly indicates that the magnification factor is quite
useful to constrain the Hubble constant even if the scatter of $\beta$
is taken into account. The contour is slightly elongated along the
$\beta$-direction by the scatter of $\beta$, but the accuracy of the
Hubble constant determination is almost same as that in Figure
\ref{fig:betah_comb}. On the other hand, the strong degeneracy still
remains when only the magnification ratio is used as constraints.
Therefore, we conclude that magnification factors which are observed
in SN Ia lensing provide indeed important information to break the
$\beta$-$h$ degeneracy. 

The most important effect we neglect here is the effect of microlensing
by stars in the lens galaxy. The magnification probability distribution
of microlensing has a long tail at high magnification regions and thus
induce non-Gaussian errors. This means that microlensing may change our
result systematically. The robust estimate of the magnification probability
distribution of microlensing is, however, difficult because the amount
of microlensing fluctuation depends on the fraction of the stellar
component which is still unclear \citep[e.g.,][]{schechter02}.
Therefore, in this paper we show the effect of microlensing by simply
adding non-Gaussian errors. We assume the following log-normal probability
distribution: 
\begin{equation}
 P(\log\mu_{\rm ML})=\frac{\exp\left\{-[\ln(\log\mu_{\rm ML}+m)]^2/2s^2\right\}}{\sqrt{2\pi s^2}(\log\mu_{\rm ML}+m)},
\end{equation} 
where $\mu_{\rm ML}$ indicates the additional magnification due to
microlensing. The parameters $m=1$ and $s=0.2$ are chosen so as to
reproduce the typical fraction which suffers from microlensing
variability; in our model the fraction magnified by $\log\mu_{\rm
ML}>0.4$ ($>0.2$) is $\sim 5$\% ($\sim 20$\%) respectively, and this
fraction is roughly consistent with the previous estimations
\citep{koopmans01,wyithe02}. The probability distribution of
microlensing should depend on convergence and shear at images and
therefore each image may have different probability distributions. We
neglect this effect because we now consider quadruple lens systems in
which images have approximately the same separations from the lens
center and are likely to have similar convergence and shear.
Constraints on $\beta$ and $h$ including the effect of microlensing are
shown in all figures (dotted lines). These figures indicate that the
effect of microlensing is fairly small. The magnification by
microlensing systematically moves contours mainly to the lower $h$, but
the deviations are sufficiently small and our assumed model ($\beta=1$
and $h=0.5$) lies still within contours. Our conclusion is therefore
that the effect of microlensing is not so severe.

\section{Summary and Discussion}\label{sec:sum}

We have shown that the strong degeneracy between radial mass profiles 
$\beta$ and the Hubble constant $h$ can be broken if we use
magnification factors as constraints. This means that SN Ia lensing has
the great advantage over traditional quasar lensing. We have found that
in the case of SN Ia lensing the Hubble constant is constrained quite
accurately, with $\sim 5$\% accuracy from only several lens events. We
have found also that both the scatter of radial mass profiles and
microlensing do not affect our result so much. In contrast to this,
quasar lensing can poorly constrain the Hubble constant, even if the
same accuracy of time delay measurements is assumed. Of course, this
method can be applicable to any distant astronomical objects which have
quite narrow luminosity function. The limitation of our method is that
it probes only the local slope of the mass profile. Therefore, the value
of the Hubble constant derived from our method may be different from the
true value if the radial mass profile is significantly different from a
power-law.   

Although SN lensing has not ever been observed, lensed SN will be
found in the future observations such as SNAP. Then how many lensed SNe
are expected to be observed in the future observational plans? SNAP survey,
for instance, can catch $2000\sim3000$ Type Ia SNe at $z\la 1.7$ per
year. Since the lensing probability at $z\sim 1.5$ is $\sim 10^{-3}$, we 
observe at least a few lensed SNe per year \citep{holz01,oguri03}. 
Time delays between images are always observed in the case of SN
lensing because SNe are transient phenomena. Hence all lensed SNe can be
used to constraint the Hubble constant. Therefore, a few year's
observation by SNAP is enough to derive the accurate value of the Hubble
constant. Note that our method can constrain the Hubble constant
accurately from even one lens system. On the other hand, large-scale
surveys, such as two degree Field system (2dF) and Sloan Digital Sky
Survey (SDSS), will also find more than one hundred of quasar lenses.
The number of lens systems for which time delays are measured, however,
should be much smaller than this because of the difficulty in measuring time
delays. Moreover, additional observations, such as host galaxies of
quasars or velocity dispersions of the lens galaxies, are needed to
break the $\beta$-$h$ degeneracy. Therefore we conclude that our method
is practically applicable and indeed have advantages compared with
methods using quasar lenses. 

In our simulation, we considered only quadruple lens systems which
have larger numbers of constraints. In practice, SNAP-like survey will
catch double lenses much more than quadruple lenses because the
magnification bias is neglected at least for low-$z$ SNe ($z\la
1.7$ in the case of SNAP survey). Time delay bias \citep{oguri03}
favors quadruple lenses, but the time delay bias is not so significant
at typical lens separation $\theta\sim 1''$. We believe, however, our
results are also applicable to double lenses, because magnification
factors are sensitive to radial mass profiles even when the source is
far from the center of the lens and thus likely to be the double lens
\citep[see][]{oguri02}. 

Finally, we comment on the measurement of the magnification factor
$\mu$. Consider the situation that only lensed SNe are observed and
the absolute magnitude of (unlensed) SNe, which is estimated using the
local Hubble constant, is known. In this case, the flux of
unlensed SNe at the same redshift, expected from the absolute magnitude, 
scales as $f_{\rm unlensed}\propto h^2$ if the assumption that the local
Hubble constant is the same as the global Hubble constant (denoted by
$h$) is relaxed. Since we observe $f_{\rm lensed}$ directly, the
magnification factor should scale as $\mu = f_{\rm lensed}/f_{\rm
unlensed} \propto h^{-2}$. This scaling is exactly same as that derived
in \S \ref{sec:ana}; this means that the $\beta$-$h$ degeneracy is never
broken in this situation. But if the redshift of lensed SN is not so
larger, huge numbers of unlensed SNe which have the similar redshift 
should be also observed. In this case, $\mu$ can be simply estimated
from the magnitude difference between lensed and unlensed SNe
independently with $h$. The only assumption needed to estimate $\mu$ is
therefore that SN Ia is an excellent standard candle. We emphasis that
several important uncertainties in the SN survey, such as cosmological
parameters, possible evolution of intrinsic luminosities, dust
extinction outside the lens galaxy, and the Cepheid calibration, do not
affect the measurement of $\mu$.  

\section*{Acknowledgments}
We thank Satoru Ikeuchi, Takahiko Matsubara, Yasushi Suto, and 
Ed Turner for useful discussions. We also thank an anonymous referee for
many useful comments.

\label{lastpage}

\end{document}